\documentclass{NOCOLS}

\title{Application of virtual seismology to DAS data in Groningen}
\author{Johno van IJsseldijk$^1$, Musab Al Hasani$^1$, Eric Verschuur$^1$, Guy Drijkoningen$^1$ and Kees Wapenaar$^1$ }
\date{\today}
\renewcommand{\organisation}{$^1$Delft University of Technology, Department of Geoscience and Engineering, Delft, The Netherlands}
\renewcommand{\runninghead}{Application of virtual seismology to DAS data in Groningen}

\renewcommand{\abstractname}{Summary}

\titlespacing*{\subsubsection}{0pt}{\baselineskip}{0pt}

\begin{document}

\begin{frontmatter}
\maketitle
\begin{abstract}
\large In this report we investigate whether and under what conditions virtual seismology via the acoustic Marchenko method can be applied to DAS data from a survey in the province of Groningen, The Netherlands. Virtual seismology allows to retrieve the band-limited Green's function between a virtual source at an arbitrary focal point in the subsurface, while accounting for all orders of multiples. The method requires the reflection response at the surface and an estimate of the traveltime between the surface and focal point. However, in order to successfully apply the method the reflection response needs to be free from surface waves and other direct waves, and properly scaled in order for the Marchenko scheme to converge. These limitations severely complicate the application of the Marchenko method to field data, especially seismic surveys on land. This report considers a full 2D geophone survey as well as a 1.5D approximation for a DAS survey, and compares the results of the virtual sources with an actual dynamite source. The results show that virtual seismology can be used to recreate the reflections recorded at the surface from the dynamite source using either geophone or DAS data.
\end{abstract}
\end{frontmatter}

\hdrsection{Introduction} 
Seismic reflection data at the Earth’s surface can be employed to create responses to virtual sources in the subsurface, observed by virtual receivers in the subsurface and physical receivers at the surface \citep{Wapenaar2018,Brackenhoff2022}. This methodology, which we call “Virtual Seismology”, is based on the 3D Marchenko method \citep{Broggini2014,Wapenaar2014}. The retrieved virtual responses consist not only of direct arrival and primary reflections, but also contain all internal multiple reflections. Virtual seismology can be used to forecast responses to induced earthquakes \citep{Brackenhoff2019b,Brackenhoff2019a}. Here we discuss our results of retrieving virtual seismic responses from reflection data obtained with an electrically driven seismic vibrator as a source and, in our case helically-wound, optical fibers with a laser interrogator (distributed acoustic sensing, DAS) as vertical-component receivers.

\begin{figure}[htbp!]
\centering
\includegraphics[width=.75\textwidth]{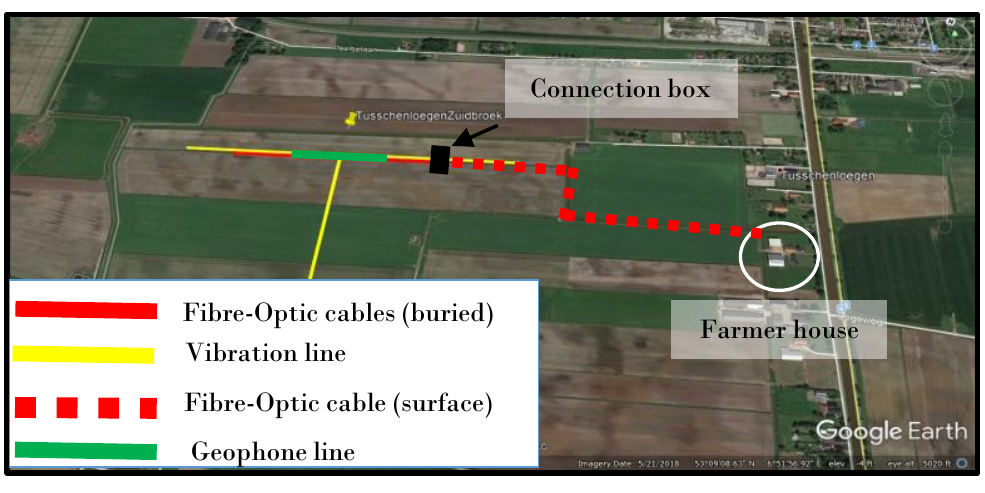}
\caption{Field map with the position of the fiber cables (surface and buried), geophone line and source line. Adapted from \citet{Hasani2023}.}
\label{fig:location}
\end{figure}

\begin{figure}[htbp!]
\centering
\includegraphics[width=.8\textwidth]{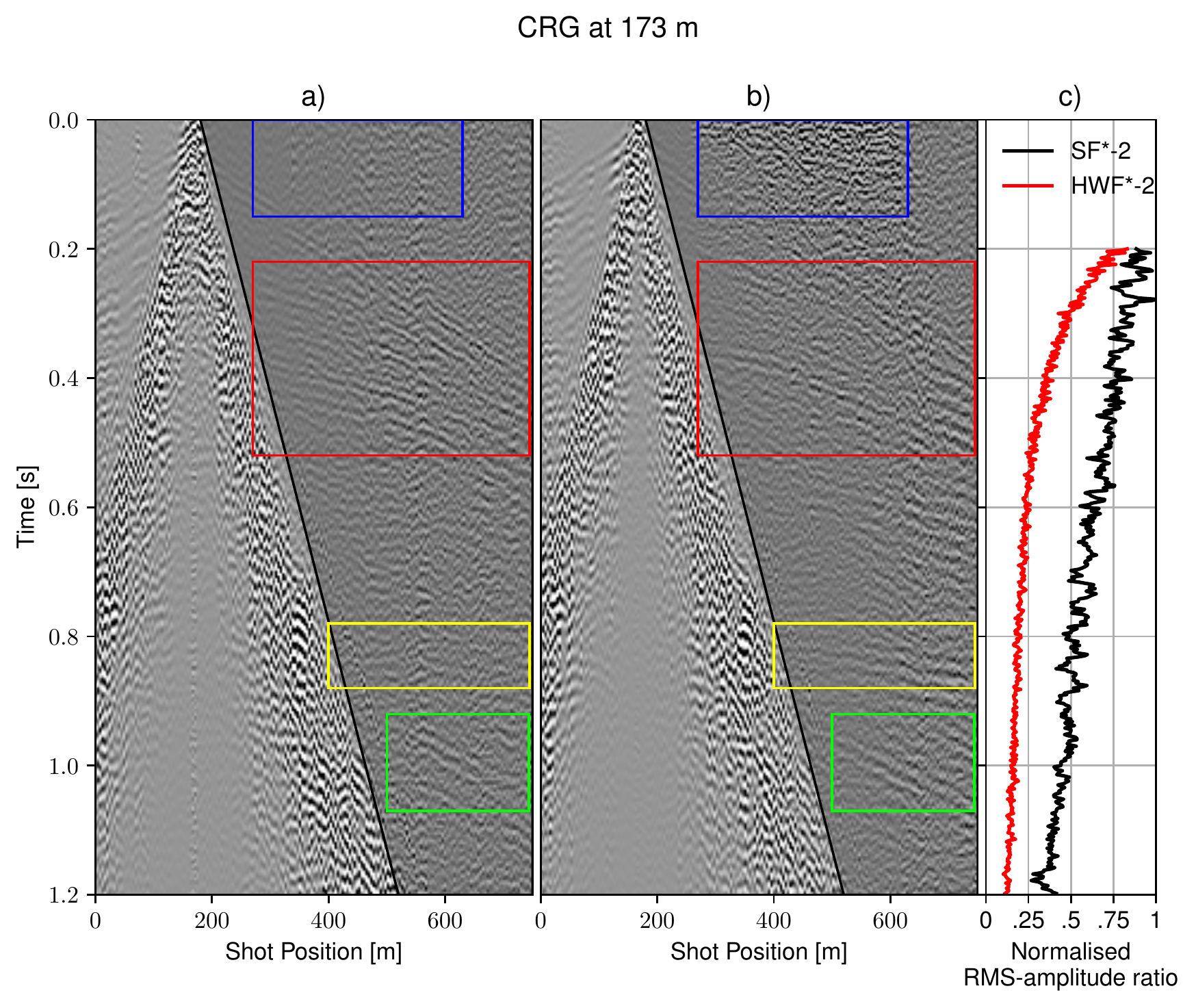}
\caption{Common-receiver gather (CRG) at 173 m of (a) straight fiber and (b) helically wound fiber, and (c) their normalized RMS values as a function of time (outside the surface-wave cones). Adapted from \citet{Hasani2023}. }
\label{fig:CRGDAS}
\end{figure}

\begin{figure}[t]
\centering
\includegraphics[width=.5\textwidth]{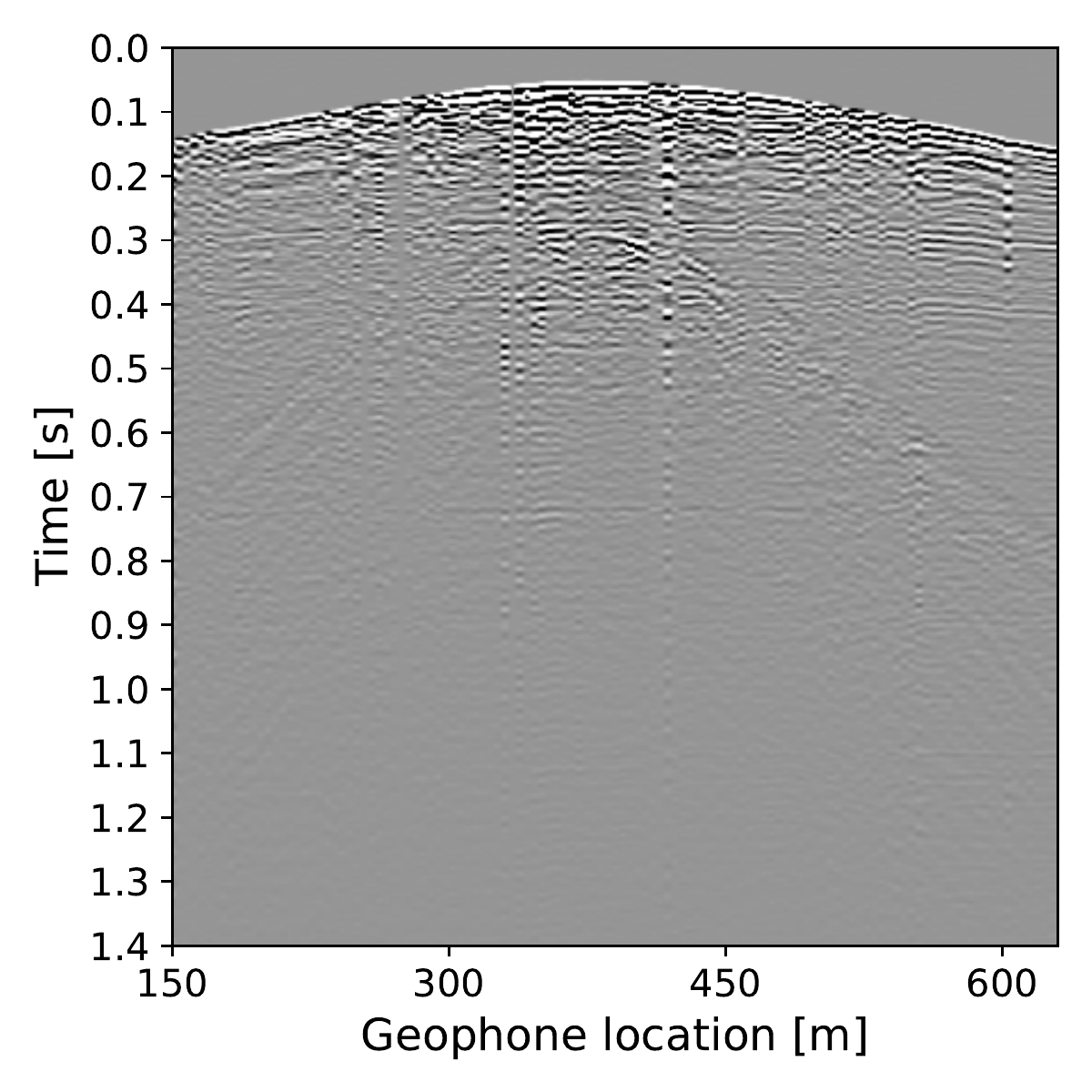}
\caption{Response to dynamite source at (x,z)=(375,100)m, registered by vertical-component geophones at the surface.}
\label{fig:Dynamite100}
\end{figure}

\hdrsection{Data acquisition}
We briefly discuss the data acquisition. For a more extensive discussion we refer to \citet{Hasani2023}. A seismic line has been deployed in the province Groningen (The Netherlands), at 53°9’16.12”N, 6°50’53.99”E (\autoref{fig:location}). An electrically driven seismic vibrator, based on linear motor technology \citep{Noorlandt2015} was used at every 2m along a source line of 750m. At each position the source was driven with a sweep signal from 2 to 180 Hz. Straight and helically wound fiber-optic cables were buried along the same line and two different types of interrogators were used to record the seismic responses at every meter with a gauge length of 2 m. Using the two types of cables allows, theoretically, retrieval of the horizontal and vertical strain rates \citep{Hasani2023}. Along a part of the line, vertical-component geophones were deployed at every 4 m, for reference. \autoref{fig:CRGDAS} shows a common-receiver gather (CRG) at 173 m, after preprocessing. The red and yellow boxes show shallow and deep P-wave reflection events, whereas the event in the green box is interpreted as a P-to-S reflection event. 
Moreover, a borehole was drilled at 375 m and dynamite sources were ignited at depths of 90, 95 and 100 m as a reference for the virtual-source responses that will be discussed later. \autoref{fig:Dynamite100} shows the response of a dynamite source at 100 m depth, registered by geophones at the surface. The goal of this work is to virtually recreate this response from the reflection data measured at the surface by means of the Marchenko method.

\hdrsection{Data pre-processing}
The raw data were correlated with the source sweep to remove the phase of the source signal from the raw data, making it zero-phase. Before the Marchenko method can be applied the reflection data need to be prepared further. This pre-processing includes removing the direct arrivals and ground roll from the data as well as applying an appropriate scaling required for the Marchenko scheme. 

\subsubsection*{Ground roll mute and interpolation}
\begin{figure}[tbp!]
\centering
\includegraphics[width=\textwidth]{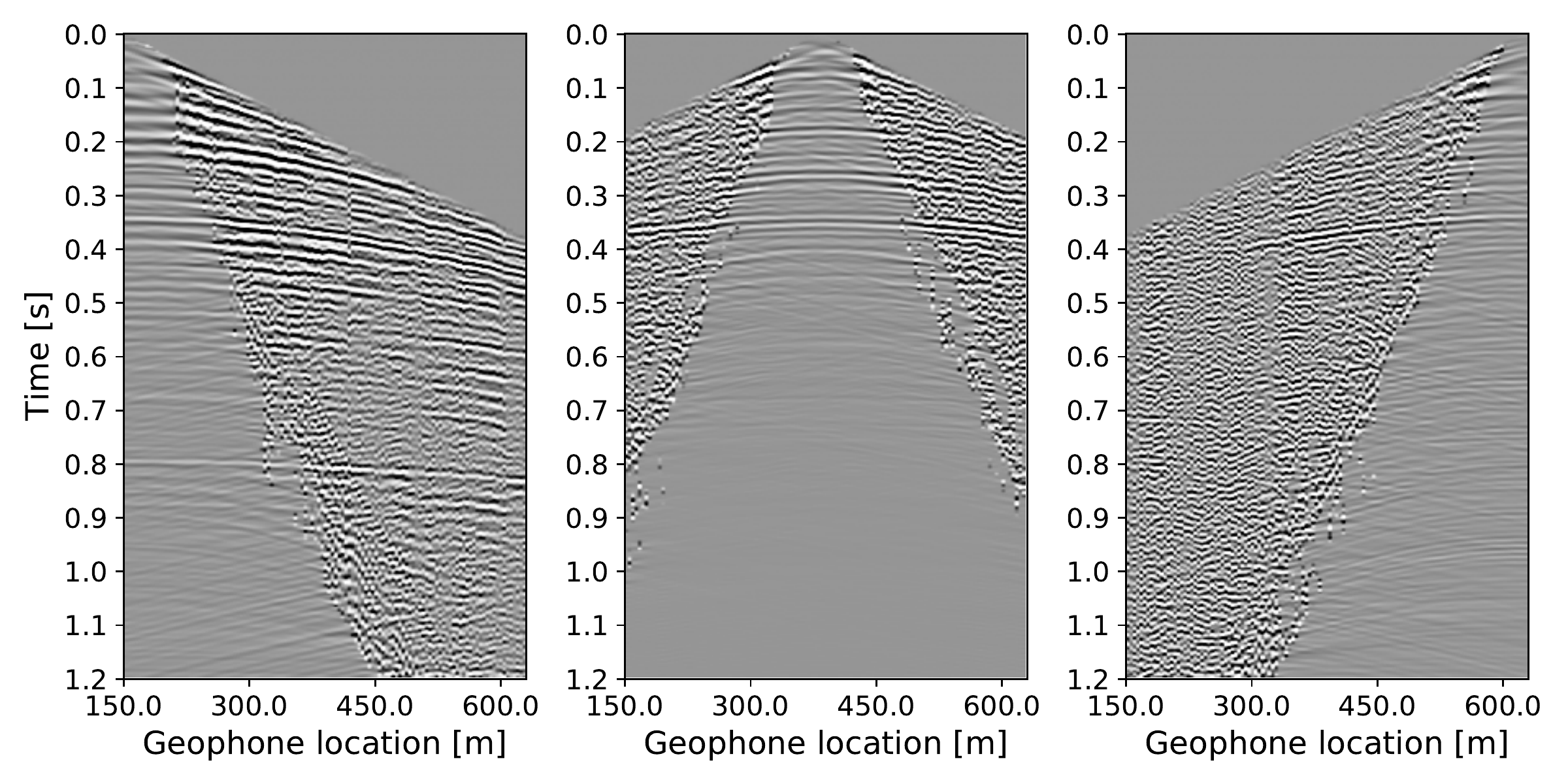}
\caption{From left to right, the first, middle and last common-source gather (CSG) at 150, 386 and 626m, respectively, on the geophone line after ground roll mute and Radon interpolation.}
\label{fig:GeophoneInterp}
\end{figure}

\begin{figure}[htbp!]
\centering
\includegraphics[width=\textwidth]{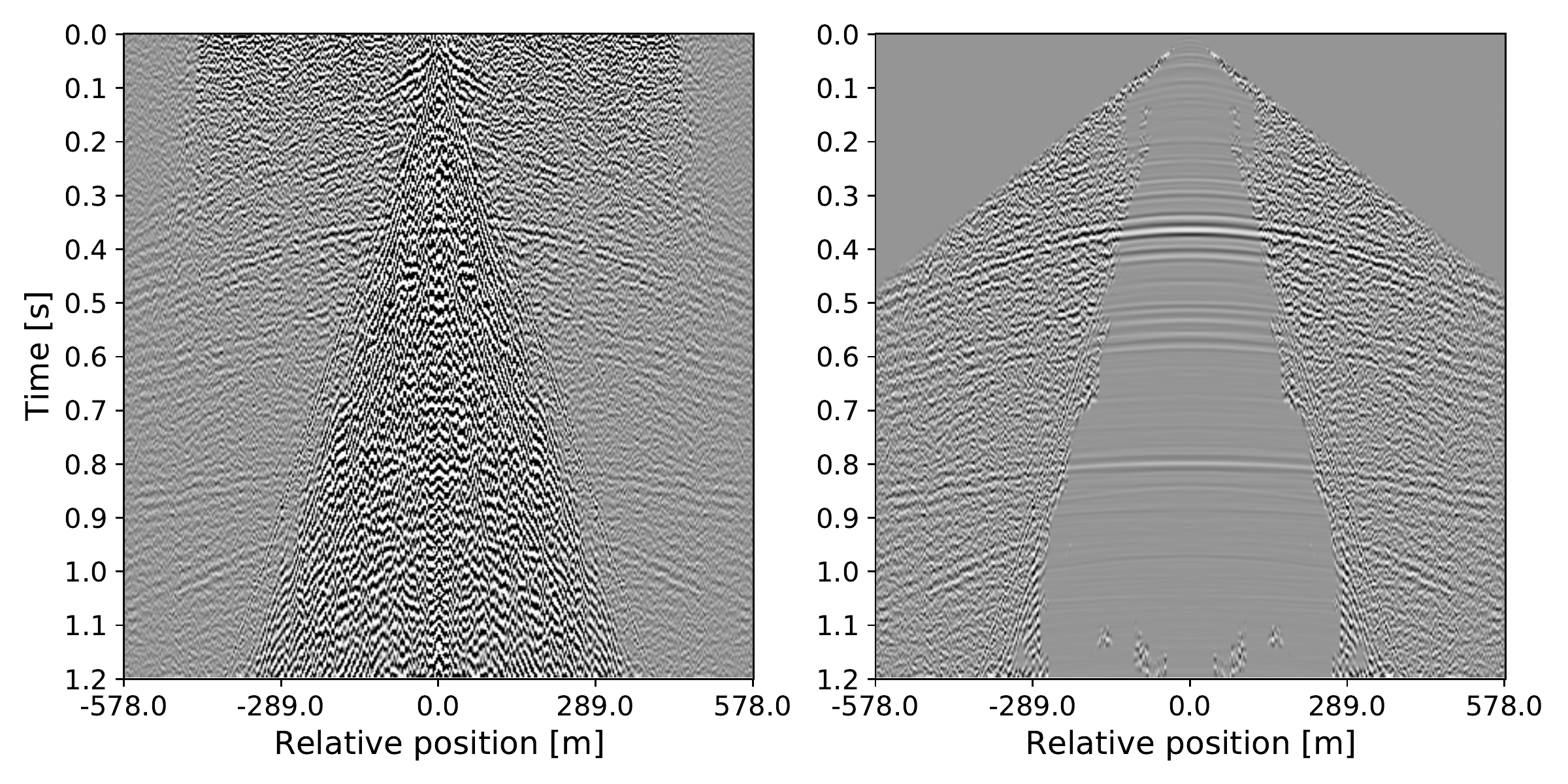}
\caption{On the left, CRG from the DAS data in \autoref{fig:CRGDAS}, the mirror image of the positive offsets of the same gather is added to create a symmetric CSG, the full 1.5D reflection response is created from this CSG. The right panel shows the same CSG after ground roll mute and Radon interpolation.}
\label{fig:DASInterp}
\end{figure}

The surface waves, as seen inside of the cone in \autoref{fig:CRGDAS}, were muted. Thereafter, normal-moveout (NMO) was applied to align the hyperbolic reflections in the data. Next, the data were transformed to the Radon domain to interpolate the muted part of the data \citep{Kabir1995}. Finally, the NMO correction is undone to obtain the interpolated data-set free from surface waves. \autoref{fig:GeophoneInterp} shows the interpolated results of three common-source gathers (CSG) on the geophone line. The first CSG in the left-most panel of \autoref{fig:GeophoneInterp} contains more reflections than can be identified in the last CSG, the right-most panel. This is likely caused due to the final shot being located on loose soil, resulting in worse coupling of the source with the earth. \\
\indent The quality of the DAS data was varying at different source locations, meaning that the Signal-to-Noise ratio (SNR) for P-wave reflections was low for a number of the CSGs. It was, therefore, decided to create a 1.5D reflection response from a single helically wound fiber CSG with clearly defined reflections instead of using all DAS CSGs. For this purpose the CRG at 173 m (\autoref{fig:CRGDAS}) was selected. The positive offsets in this gather were mirrored to create the symmetric gather as shown in \autoref{fig:DASInterp}. Once again, the ground roll is muted and interpolated using the Radon transform on the NMO-corrected CSG. After undoing the NMO, the interpolated gather is shown on the right in \autoref{fig:DASInterp}. From this gather a full line with 289 shots and a fixed spread of 289 receivers is produced, that is a 578 by 578 m reflection response. Since the single CSG is used for all source positions, the reflector at 0.8 s as well as the P-to-S reflection will be clearly visible in all CSGs, contrary to the CSGs of the geophone line that differ in SNR between different shot locations.

\subsubsection*{Scaling correction of the reflection data}
In order for the Marchenko method to converge, the amplitude of the reflection data have to be properly scaled. Firstly, to correct for 3D geometrical spreading (approximately $1/t$) on the 2D ($\approx 1/\sqrt{t}$) scheme a factor of $\sqrt{t}$ is applied to the data. Next, we correct for absorption and other effects that are for example related to the source signature or the interpolation. This is achieved by minimizing the cost functions as described in \citet{Brackenhoff2016}. Different gains and linear factors are considered to find an optimal factor of the form $a e^{bt}$, where $a$ is a linear, multiplicative factor and $bt$ a time-dependent exponential gain. However, there are other factors, such as amplitude versus offset behaviour, that have not been considered, meaning that the final scaling factor may not give the optimal results. The consequences and limitations of these decisions will be discussed at the end of this report. 

\begin{figure}[htbp!]
\centering
\includegraphics[width=\textwidth]{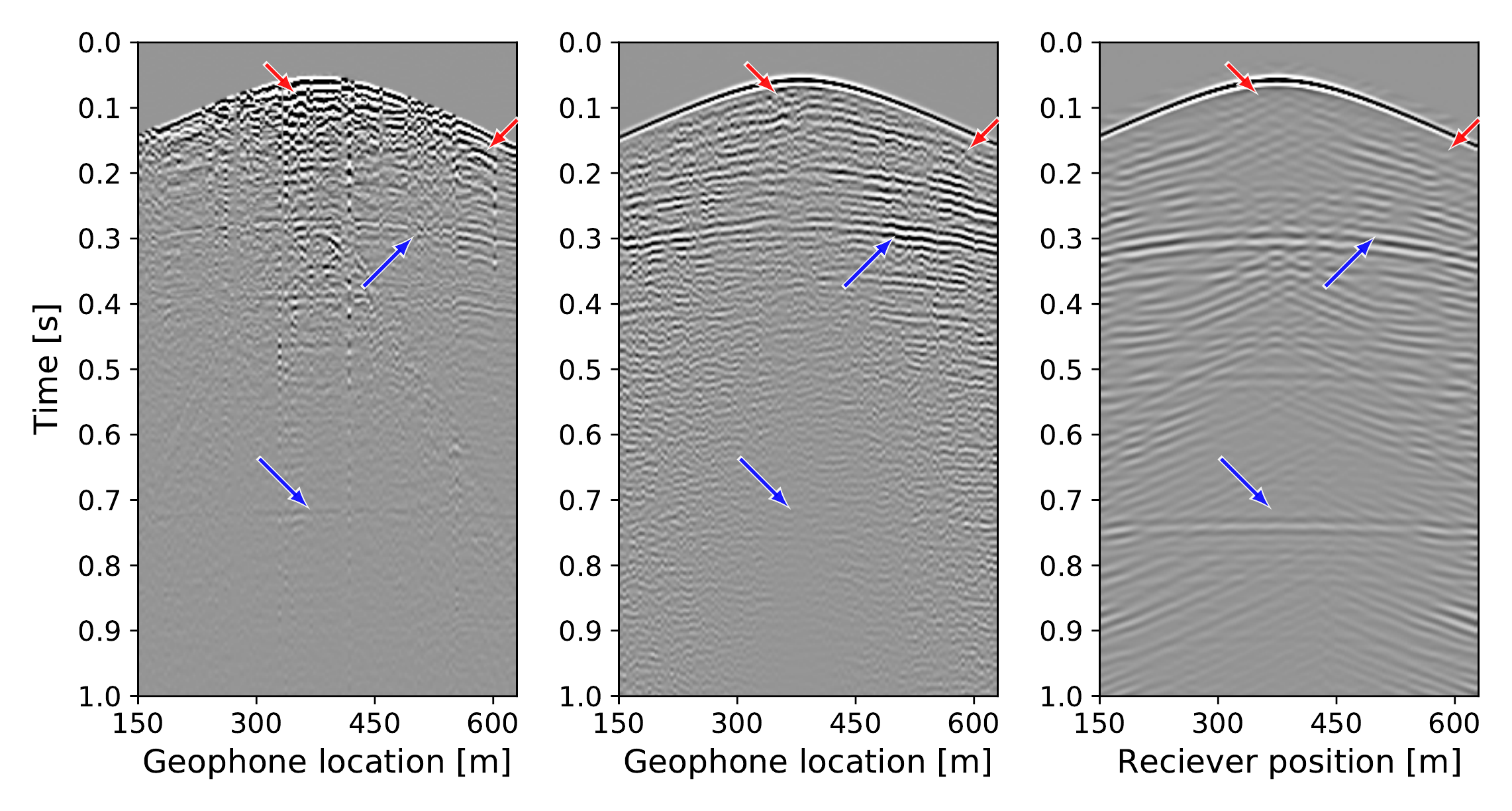}
\caption{From left to right: reference dynamite shot at 100 m depth (same as \autoref{fig:Dynamite100}), virtual shot at 100 m from the geophone data and the same virtual shot acquired with the DAS data. The blue arrows indicate the first and second primary event.}
\label{fig:Results}	
\end{figure}

\hdrsection{Application of virtual seismology}
The next step before applying virtual seismology is finding an estimate of the traveltime between the focal point and the surface. Here, this is achieved by acoustic finite-difference modeling the acoustic Green's function between the receivers on the surface and the focal point at x = 375 m and a depth of 100 m (i.e. at the same location as the dynamite source). Subsequently, the first arrival of this Green's function is time-reversed, and used as initial estimation of the focusing function for the Marchenko method. The iterative Marchenko scheme is then used to find the upgoing- and downgoing-Green's functions between the focal point and the surface \citep{Thorbecke2017}. These decomposed Green's functions are summed to find the two-way acoustic Green's function, which can then be compared to the response of the dynamite shot at depth, shown on the left of \autoref{fig:Results}. The results of 8 iterations applied to the geophone data are shown in the middle panel of this figure. The reflection at 0.3 s, marked by a blue arrow, is clearly visible in both gathers. Furthermore, the red arrows show some similarities after the direct arrival of the acoustic Green's function. The rightmost gather of \autoref{fig:Results} shows the results of applying the Marchenko scheme to the DAS data. Now the second reflection at approximately 0.8s, marked by a second blue arrow, also becomes visible in the data, contrary to the geophone results. This result is explained by the fact that the 1.5D DAS data contains two obvious reflections at 0.4 s and 0.8 s, whereas this second reflection at 0.8 s is not always visible in all the geophone CSGs. The downside of using a 1.5D approximation for the DAS data is that it is unable to resolve finer details, which is especially visible after the direct arrival (red arrows); the geophone results are able to resolve some heterogeneity in the signal here, whereas the DAS results display no similarities with the dynamite shot at the same traveltimes. \pagebreak

\hdrsection{Discussion and conclusions}
In this report we assessed the feasibility of applying our virtual seismology methodology to seismic land datasets of geophone as well as DAS recordings. A number of challenges need to be overcome to properly apply the method. Firstly, the direct and surface waves need to be removed from the data, which was achieved by applying a strict mute to eliminate the ground roll and interpolating the subsequent gap in the Radon domain. The second challenge is finding the correct scaling of the reflection response, in order for the Marchenko scheme to converge to a solution, and ensure that internal multiples are properly accounted for. \\
\indent The surface waves, as well as the direct P-waves, are easily removed by the mute, and the interpolation was able to reconstruct the hyperbolic reflections in the data. The downside of this method is that near-surface reflections that mostly overlap with the ground roll will be destroyed in the process. This means that a valuable part of the data is lost, and internal multiples from these reflectors can no longer be resolved with the Marchenko method. Additionally, the refracted waves will be affected, this is of lesser concern as the current Marchenko method does not consider refractions. Future research should consider how these reflections can be preserved or recreated from their multiple reflections, especially because the near-surface can be a prolific source of internal multiples. A secondary consequence of the Radon interpolation is the inaccurate amplitudes of the interpolated reflections, we hope to compensate for this effect by applying the scaling factors. \\
\indent The scaling factors are designed to compensate for a number of factors that reduce the accuracy of the amplitude in the data. Nevertheless they are far from perfect, and aside from the previously mentioned loss of accuracy due to the interpolation, a number of additional parameters need to be considered. Firstly, the geometrical spreading estimation is not fully accurate, as \citet{Dukalski2022} show. Another scaling error is introduced by the fact that the formulation of the Marchenko equations assumes a dipole source ($F_z$) with monopole receivers ($P$), or via reciprocity a monopole source ($Q$) with dipole receiver ($v_z$). On the contrary, our survey has both a dipole source as well as dipole receivers. Moreover, wavefield damping and imperfect source-phase removal are other causes of erroneous amplitudes in the reflection data. On the one hand, all these factors are mainly linear or time-dependent, and can, therefore, be corrected using a formulation with some linear and some time-dependent scaling factors. On the other hand, erroneous amplitude versus offset effects are not accounted for in our formulation, thus they are likely to cause bigger issues. These changes become especially relevant when considering DAS data, as the helically wound fiber-optic lines measure a combination of vertical and horizontal strain rather than the vertical particle velocity measured by the geophones. While the vertical strain can, in theory, be retrieved using both the helically wound and straight fiber recordings, the exact relation between these two measurements remains uncertain \citep{Hasani2023}. Moreover, the resulting strain-rate would still contain a spatial derivative compared to the vertical particle velocity measured by the geophones. Ultimately, future research is required to more carefully study the effects of different scaling errors in the data, and how to properly correct for each one of them. \\
\indent One of the issues that has not been discussed before is the error due to the approximation of the initial focusing function. Specifically, forward-scattering, due to diffractions or other sharp discontinuities, means that estimating the initial focusing function with the direct P-wave arrival of the Green's function is no longer accurate. These effects can be seen in the first event of the dynamite source (\autoref{fig:Dynamite100}), which is not continuous everywhere. \citet{vanderNeut2022} propose an adaption to the Marchenko equations with transmission data to correct for these deviations. In theory, a similar approach using the direct arrival of the dynamite shot can be used to update the estimated Green's functions in this study. However, the method does not take into account velocities contrasts, hence the application to the current study remains challenging. \\
\indent Next, the current formulation does not take into account free-surface multiples, because the subsurface is relatively unconsolidated and some weak surface-related multiples are expected. However, in the case of seismic surveys on a more solid underground these multiples will become more obvious. In this case a formulation that includes free-surface multiples might be required \citep{Singh2017}, or alternatively the free-surface multiples have to be removed before applying the method by means of for example surface-related multiple elimination \citep{Verschuur1992}. \\
\indent A final concern is the fact that the Marchenko method that was used is designed for acoustic wavefields, whereas the data was recorded in an elastic medium. \citet{Reinicke2021} conclude that while elastic effects have a substantial impact on the results of the Marchenko method, their impact is limited and unlikely to alter interpretation of migrated sections. Moreover, current research is aimed at developing elastic adaptations of the Marchenko method \citep{Reinicke2019}. \\
\indent To conclude, we have found that the virtual seismology methodology can be used to redatum wavefields in the subsurface, and virtually estimate the response to a dynamite source at depth, observed at the surface. This estimation is achieved solely with the reflection response at the surface, and an estimation of the direct arrival of the acoustic Green's function between the focal depth and the surface. It was shown that this was possible using geophone as well as DAS data. Future research should mainly focus on how the data can be scaled in order for the Marchenko scheme to properly converge and find an appropriate solution that includes all internal multiples. Such research should examine how to deal with linear (i.e. multiplicative), time-dependent and offset-dependent scaling errors in the reflection data. Additionally, the results can be further improved if the near-surface reflections can be resolved better, as a majority of the internal multiples are oftentimes generated here.

\subsection*{Acknowledgement}
This research was funded by the European Union's Horizon 2020 research and innovation programme: European Research Council (grant agreement 742703).

\setlength{\bibsep}{0pt plus 0.3ex}
\bibliography{Groningen}

\end{document}